\documentclass[aps,prl,twocolumn,superscriptaddress,showpacs]{revtex4}
\usepackage{epsfig}
\usepackage{amsmath}
\usepackage{multirow}%
\usepackage{array}%
\usepackage[dvipsnames,usenames]{color}

\usepackage{epstopdf}
\begin{document}

\title{Evolution of topological end states in the one-dimensional Kondo-Heisenberg model with site modulation}

\author{Neng Xie}
\author{Danqing Hu}
\affiliation{Beijing National Laboratory for Condensed Matter Physics, and Institute of Physics,
Chinese Academy of Sciences, Beijing 100190, China}
\author{Shu Chen}
\author{Yi-feng Yang}
\email[]{yifeng@iphy.ac.cn}
\affiliation{Beijing National Laboratory for Condensed Matter Physics and Institute of Physics,
Chinese Academy of Sciences, Beijing 100190, China}
\affiliation{School of Physical Sciences, University of Chinese Academy of Sciences,
Beijing 100190, China}
\affiliation{Songshan Lake Materials Laboratory, Dongguan, Guangdong 523808, China}

\date{\today}

\begin{abstract}
We investigate the interplay of the topological and Kondo effects in a one-dimensional Kondo-Heisenberg model with nontrivial conduction band using the density matrix renormalization group method. By analyzing the density profile, the local hybridization, and the spin/charge gap, we find that the Kondo effect can be destructed at the edges of the chain by the topological end state below a finite critical Kondo coupling $J_{K}^{c}$. We construct a phase diagram characterizing the transition of the end states.
\end{abstract}


\maketitle

Topological quantum states of matter as revealed in the tremendous studies of topological insulators have attracted intensive interests in recent years \cite{Hasan2010,Qi2011}. In correspondence to the bulk insulating state is an exotic metallic state with spin-momentum locking on the surface \cite{Kane2005}. This special surface state stems from the nontrivial topology of the electronic band structures in the parameter space and is robust against weak disorder and impurity scattering \cite{Bernevig2005}. Topological nontrivial phases and the corresponding end states can also be realized in the one-dimensional quasiperiodic optical lattice through periodical modulation of additional parameter as a new dimension besides momentum \cite{Lang2012, Kraus2012}. This is intrinsically connected to the two-dimensional Hofstadter lattices with quantum Hall effect \cite{Hofstadter1976} and can be detected by the density profile.

Electronic correlations may interact with topological properties to yield more exotic many-body quantum phenomena. One interesting example is the interplay of the topological state and the Kondo physics, giving rise to the so-called topological Kondo semimetal/insulator \cite{Dzero2010, Lu2013, Wolgast2013, Kim2014, Zhang2018CPB}, its breakdown on the surface \cite{Alexandrov2015,Erten2015}, the emergence of new topological insulator phase in the Kondo-screened case \cite{Feng2011}, and possible topological phase transition with pressure \cite{Zhou2016}. The tractable optical lattice systems in one dimension (1D) provide a unique experimental and theoretical platform to simulate their interplay. It should be noted that our model is different from that of topological Kondo insulators, but may well be realized in real materials or heterostructures with Kondo-coupled magnetic and topological layers.

In this work, we consider half-filled trapped fermions on a 1D lattice with nontrivial topological properties coupled to a background Heisenberg spin chain by local Kondo interactions. We adopt the exact density matrix renormalization group method (DMRG) for the numerical simulation of the ground state properties \cite{White1992,White1993,Schollwock2005}. This allows us to analyze the density profile, the local hybridization, the spin/charge gap, and their evolution with external parameters and the strength of the local Kondo coupling. We find a phase transition from the topological state to the Kondo singlet state at a finite critical Kondo coupling. At the edges of the chain, we find  that the conduction electrons are modified, causing a suppression of the Kondo effect below the critical coupling. Our major conclusions may still hold in higher-dimensional systems.

\begin{figure}[t]
\begin{center}
\includegraphics[width=0.45\textwidth]{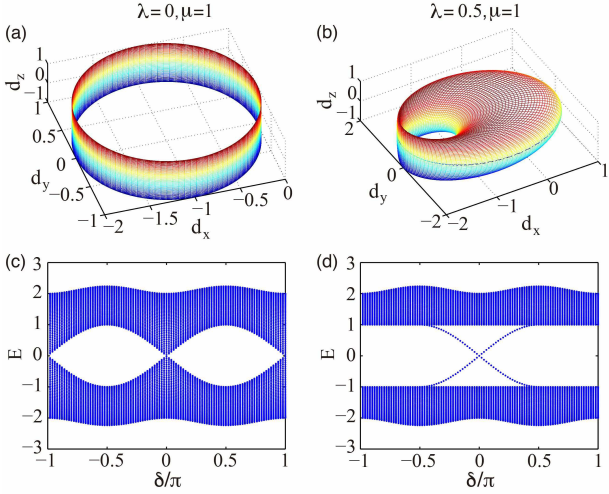}
\caption{(Color online) (a) and (b) Geometry structures for $\vec{d}(k)$ with $\mu=1$ and $\lambda=0$, 0.5. (c) and (d) The energy spectra for $\lambda=0$ (c) and 0.5 (d) from Eq. (\ref{eig}). The chemical potential is set to $\mu=1$ and the lattice site is $N=40$ with open boundary conditions.}
\label{fig1}
\end{center}
\end{figure}

We consider the following model Hamiltonian:
\begin{eqnarray}
H&=&\sum^{N-1}_{i=1}\sum_{\sigma=\pm}(t_{i}c_{i,\sigma}^{\dagger}c_{i+1,\sigma}+\text{H.c.})+\sum^{N}_{i=1}\sum_{\sigma=\pm}\mu_i c_{i,\sigma}^{\dagger}c_{i,\sigma}\nonumber\\
&&+ J_K\sum^{N}_{i=1}\vec{S}_i\cdot\vec{s}_i + J_H\sum^{N-1}_{i=1}\vec{S}_i\cdot\vec{S}_{i+1},
\label{Ham}
\end{eqnarray}
where $c_{i,\sigma}^{\dagger }$($c_{i,\sigma}$) creates (annihilates) a conduction electron with spin $\sigma$ at the $i$-th site, and $\vec{S}_{i}$ is the $S=1/2$ spin operator of the localized spins. The spin density operators of the conduction electrons are $\vec{s}_{i}=\sum_{\alpha,\beta}c_{i,\alpha}^{\dagger }(\vec{\sigma}/2)_{\alpha\beta} c_{i,\beta}$, where $\vec{\sigma}$ are the Pauli matrices. We choose $J_K>0$ for an antiferromagnetic Kondo coupling. For DMRG calculations, we use a modified {\scriptsize{DMRG++}} code \cite{Alvarez2009,Xie2015PRB,Xie2017SP}. The phase space is restricted by the two good quantum numbers: 
$S^z=\sum_i s_i^z+ \sum_i S_i^z$ and $N^c=\sum_{i\sigma}c_{i\sigma}^\dagger c_{i\sigma}$. To simulate the topological quantum state, the 1D optical lattice requires a special modulation on the hopping term, $t_{i}=1+(-1)^{i}\lambda\cos\delta$, and the chemical potential, $\mu_i=(-1)^{i}\mu\sin\delta$. The latter represents the dimerization strength with $\delta$ as the cyclical modulate parameter varying from $0$ to $2\pi$. As illustrated in Fig.~\ref{fig1}(a), this gives two sublattices (denoted as $A$ and $B$) with different chemical potentials, $\mu_{A}=-\mu \sin\delta$ and $\mu_{B}=\mu \sin\delta$, and the inter-sublattice hopping, $t_1=1-\lambda\cos\delta$ and $t_2=1+\lambda\cos\delta$. This model is also known as the Rice-Mele model if $\delta=\omega t$ \cite{Rice1982}. At $\mu=0$, it reduces to the Su-Schrieffer-Heeger (SSH) model \cite{Su1979}. For periodic boundary conditions, performing the Fourier transformation, $c_{A/B,j}=\sqrt{2/N}\sum_{k}e^{i k j}c_{A/B,k}$, we have the free Hamiltonian in a two-level form in the momentum space,
\begin{eqnarray}
H_c&=&\psi_{k}^{\dagger}\begin{pmatrix}
\mu_A & t_1 + t_2 e^{-ik}\\ 
t_1 + t_2 e^{ik} & \mu_B
\end{pmatrix}\psi_{k},\nonumber\\
&=&\psi_{k}^{\dagger}\left[\vec{d}(k)\cdot\vec{\sigma}\right]\psi_{k},
\end{eqnarray}
where  $\psi_{k}^{\dagger}=(c_{A,k}^{\dagger},c_{B,k}^{\dagger})$ and $\vec{\sigma}$ are the Pauli matrices acting on the pseudospin $\psi_{k}$. We have $d_x(k)=t_1+t_2\cos k$, $d_y(k)=t_2\sin k$ and $d_z(k)=-\mu\sin\delta$.

The topological property of the free Hamiltonian can be seen from the the solid angle swept out by $\vec{d}(k)$ \cite{Xiao2010}. As shown in Fig.~\ref{fig1}(b) for $\mu=1$, $\vec{d}(k)$ forms a closed surface topologically equivalent to the sphere $S^2$. For $\lambda=0.5$, the surface contains the original point $\vec{d}=0$, thus contributing a finite solid angle, whereas for $\lambda=0$, the original point locates on the surface so that the solid angle is zero. We further calculate their respective Chern number numerically in the parameter space $(k,\delta)$ \cite{Fukui2005}:
\begin{equation}
\begin{aligned}
c_{n}=\frac{1}{2\pi}\int_{0}^{2\pi}d\delta\int_{0}^{2\pi}dk ( \partial_{\delta}A_{k}-\partial_{k}A_{\delta}),
\end{aligned}
\end{equation}
where $A_{k/\delta}=i\left \langle \phi _{n}(k,\delta)\right | \partial_{k/\delta} \left | \phi _{n}(k,\delta)  \right \rangle $ is the Berry connection and $\left | \phi _{n}(k,\delta) \right \rangle$ is the occupied Bloch state of the $n$-th energy band. We obtain $c_1=1$ and $c_2=-1$ for $\lambda=0.5$ and $c_{1/2}=0$ for $\lambda=0$. This proves that the modulated 1D optical lattice with free conduction electrons can host a topologically nontrivial state at half filling for $\mu=1$ and $\lambda=0.5$.

\begin{figure}[t]
\begin{center}
\includegraphics[width=0.5\textwidth]{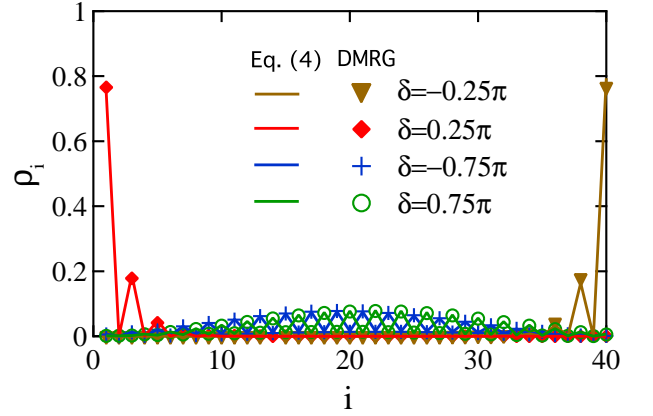}
\caption{(Color online) Comparison of the density distribution of the end states between analytical and DMRG calculations with open boundary conditions. The parameters are $N=40$, $\mu=1$, and $\lambda=0.5$.}
\label{fig2}
\end{center}
\end{figure}

To see this more clearly, we write out explicitly the single particle eigenstates, $\left | \psi_n \right  \rangle=\sum_i u_{i,m} c^{\dagger}_i \left | 0 \right \rangle$ with $H_c\left | \psi_n \right\rangle=E_n\left | \psi_n \right\rangle$, and evaluate the eigenvalue equation,
\begin{equation}
t_{i}u_{i+1,n}+t_{i-1}u_{i-1,n}+\mu_{i}u_{i,n}=E_{n}u_{i,n}.
\label{eig}
\end{equation}
Figures~\ref{fig1}(c) and \ref{fig1}(d) compares the energy spectra for $\lambda=0$ and 0.5. For $\lambda=0$, the band gap is closed at $\delta=\pm\pi,0$, whereas for $\lambda=0.5$, we see two lines connecting the valence and conduction bands which correspond to the two end states (or edge states in 2D) under open boundary conditions. They reduce to the two-fold degenerate zero-mode in the SSH model ($\mu=0$) owing to the particle-hole symmetry and the inversion symmetry \cite{Su1979}. A finite $\mu$ tunes the zero-mode into two end states that cross at $\delta=0$. The existence of the end states is also reflected in the density difference, $\rho_{i}=[n^c_i(N^c=N)-n^c_i(N^{c}=N-2)]/2$, which subtracts a topologically trivial background and singles out the topological end states that are unoccupied for $N^c=N-2$. Figure~\ref{fig2} compares the results derived from the eigenvalue equation and DMRG calculations. For $\delta=\pm 0.25\pi$, we see a large $\rho_i$ at one end of the chain which decays rapidly into the bulk, indicating that the extra electrons accumulate on the boundary to form the end state, whereas for $\delta=\pm 0.75\pi$, the extra electrons spread all over the bulk and no discernible accumulation is seen at the ends. This $\delta$-dependence is consistent with the energy spectra in Fig.~\ref{fig1}(c). The good agreement justifies that we may use the local occupation number $n^c_i$ in the DMRG calculations to detect the topological states.

Now we introduce the Kondo coupling with the Heisenberg spin chain and discuss the interplay between the topological state and the many-body Kondo correlations for $\mu=1$ and $\lambda=0.5$. We restrict ourselves to $S^z=0$ (nonmagnetic) and $N^c/N=1$ (half-filling). The exchange coupling between the nearest-neighbor Heisenberg spins is set to $J_H/t=0.5$ to provide background antiferromagnetic spin fluctuations. The DMRG calculations use $400$ block states for $N=40$ sites and open boundary conditions. The results have been verified to be convergent with different chain lengths and block states.

\begin{figure}[t]
\begin{center}
\includegraphics[width=0.45\textwidth]{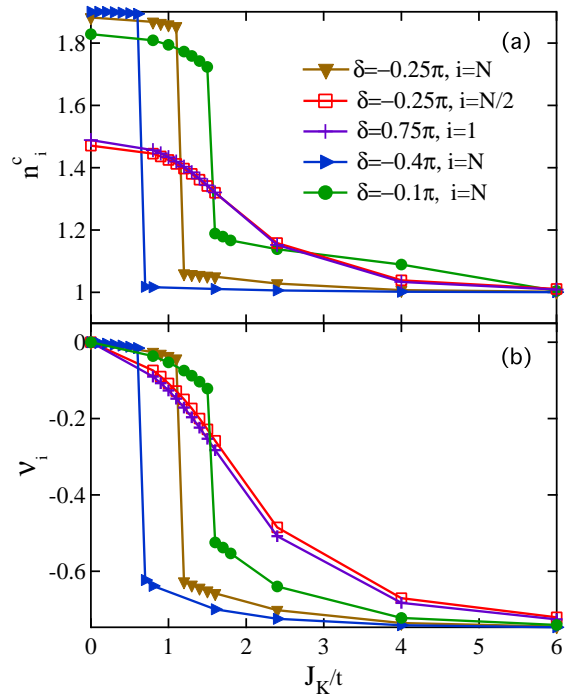}
\caption{(Color online) The density distribution and local hybridization as a function of $J_K/t$ for different phase $\delta$ on the lattice sites $i$ near the edge with a larger $n_i^c$. We set $n^c=1$, $N=40$, $\mu=1$, and $\lambda=0.5$.}
\label{fig3}
\end{center}
\end{figure}

Figure~\ref{fig3} plots the calculated local charge density, $n^c_i$, and the local hybridization, $V_i=\langle \vec{S}_i\cdot\vec{s}_i\rangle$, for varying $J_K$ and $\delta$. For $J_K=0$, the resulting $n^c_{i=N}$ at $\delta=-0.25\pi$, $-0.4\pi$ and $-0.1\pi$ are all close to 1.9, whereas $n^c_{i=N/2}$ at $\delta=-0.25\pi$ in the bulk on the same sublattice is only about 1.5 (due to finite $\mu=1$). This large difference cannot originate from the boundary effects. To see this, we take a reference system with $\delta=0.75\pi$, which corresponds to the spatial inversion of $\delta=-0.25\pi$ but has no end state. The resulting $n^c_{i=1}$ at $\delta=0.75\pi$ is plotted for comparison and found to follow closely with $n^c_{i=N/2}$ at $\delta=-0.25\pi$ for all $J_K$. This excludes the boundary effects and confirms that the nearly double occupancy at $i=N$ for small negative $\delta$ has a topological origin, in good agreement with that seen in Fig.~\ref{fig1}(c) and Fig.~\ref{fig2}. The $J_K=0$ results provide a consistency check for our following discussions at finite $J_K$.

\begin{figure}[t]
\begin{center}
\includegraphics[width=0.45\textwidth]{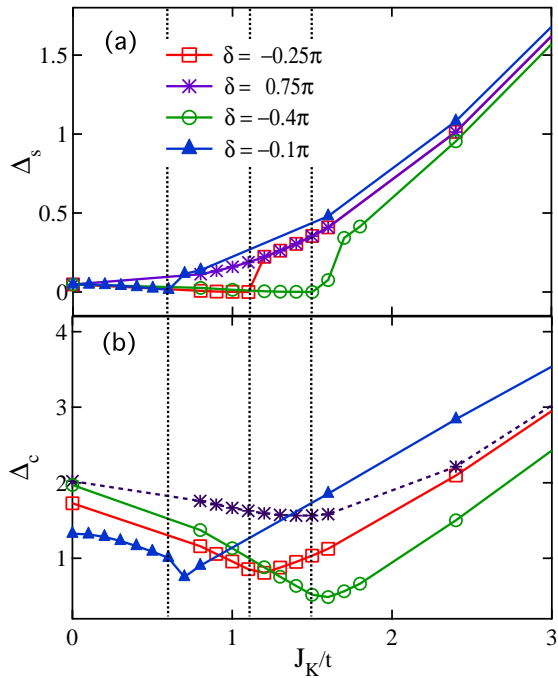}
\caption{(Color online) Evolution of (a) the spin gap and (b) the charge gap with varying $J_K/t$ for different phase $\delta$. The dash dotted line marks the results at the critical Kondo coupling, $J_K^c$. Other parameters are $n^c=1$, $N=40$, $\mu=1$, and $\lambda=0.5$.}
\label{fig4}
\end{center}
\end{figure} 

Switching on the Kondo interaction gives rise to a finite hybridization $V_i$ whose magnitude increases with increasing $J_K$. On the other hand, the local occupation numbers of the conduction electrons on the two sublattices tend to be averaged and both approach unity (half-filling) for sufficiently large $J_K$, indicating that the Kondo energy is playing a role and attempting to overcome the difference in the local chemical potentials in order to balance the spin screening on both sublattices. For the bulk at $i=N/2$, $n^c_{i}$ and $V_{i}$ change smoothly with increasing $J_K$, which is typically expected and of no surprise. However, for the end of the chain at $i=N$, the two quantities exhibit a simultaneously sharp change at a finite $J_K^c$. This feature is not seen for $\delta=0.75\pi$ and represents a property of the topological end state in competition with the Kondo interaction. For $J_K<J_K^c$, the magnitude of $V_{i=N}$ increases only slightly with increasing $J_K$ and $n^c_{i=N}$ remains large compared to its bulk value, indicating that the end state is robust against the Kondo interaction. As a result, the Kondo screening is weakened on the boundary and becomes less effective compared to the bulk. However, for $J_K>J_K^c$, the system seems to become  more strongly hybridized and, instead of recovering their bulk values, both quantities jump even closer to their strong coupling limit. Taking $\delta=-0.25\pi$ as an example, the obtained $n^c_{i=N}$ approaches unity, indicating that there is no longer topological end state at $i=N$. Simultaneously, $V_{i=N}$ jumps to almost -0.6, close to the value (-3/4) for a fully formed spin singlet. This indicates that there exists only one electron at $i=N$ which couples strongly to the local Heisenberg spin. In contrast, the bulk electrons on the same sublattice are still weakly hybridized with $V_{i=N/2}\approx -0.15$.

The anomalous weakening/enhancement of the hybridization for $J_K>J_K^c$ indicates that the end has a very different Kondo physics compared to the bulk. To see this more clearly, we calculate the charge and spin gaps of the entire lattice system,
\begin{eqnarray}
\Delta_c&=&\frac{E_0(N+2,0)+E_0(N-2,0)}{2}-E_0(N,0),\nonumber\\
\Delta_s&=&E_0(N,1)-E_0(N,0),
\label{gap}
\end{eqnarray}
where $E_0(N,0)$ is the ground state energy for total $N^c=N$ and $S^z=0$ with the chosen parameters. $E_0(N\pm2,0)$ are the excited state energy for total $N^c=N\pm2$ and $S^z=0$, and $E_0(N,1)$ is the excited state energy for total $N^c=N$ and $S^z=1$. The spin gap $\Delta_s$ hence measures the energy increase from $S^z=0$ to $S^z=1$ and $\Delta_c$ measures the average energy required to add or remove two conduction electrons from the system. Note that $N\pm2$ are used to avoid the influence of magnetic energies since $N\pm 1$ both are odd and have magnetic ground states. Since $\Delta_{s/c}$ measure the minimal energy costs for the spin/charge excitations, their behavior provides a direct comparison between the energy gap for end and bulk excitations. 

\begin{figure}[t]
\begin{center}
\includegraphics[width=0.5\textwidth]{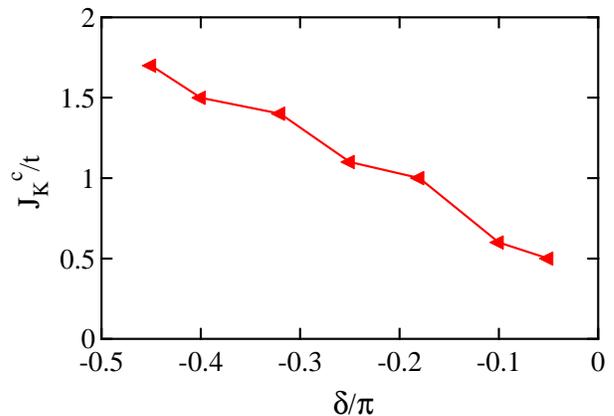}
\caption{(Color online) Evolution of the critical Kondo coupling $J_K^c$ as a function of $\delta$, marking the transition from the topological state to the Kondo state at the edges of the chain. The parameters are $n^c=1$, $N=40$, $\mu=1$, and $\lambda=0.5$.}
\label{fig5}
\end{center}
\end{figure}

As shown in Fig.~\ref{fig4}(a), we see a continuous growth of $\Delta_s$ in the bulk ($\delta=0.75\pi$) with increasing $J_K$, but a sudden jump at $\delta=-0.25\pi$ from nearly zero to the bulk value at $J_K^c$. Hence for $J_K<J_K^c$, the electrons at the end are effectively decoupled from the Heisenberg spin and have a smaller spin gap due to their special topological property, whereas for $J_K>J_K^c$, the topological effect is destroyed and, accompanying with the dissipation of the accumulated electrons, a local spin singlet is formed at the end, which gives rise to a larger spin gap than that of the bulk. In both cases, the end state seems to be detached from the bulk. 

In contrast to the sudden jump in $\Delta_s$, the charge gap $\Delta_c$ shows no discontinuity but a minimum at $J_K^c$. For $J_K=0$, as shown in Fig.~\ref{fig1}(c), $\Delta_c$ increases with increasing $|\delta|$ and measures the sum of the overall band gap, $2\lambda$, and the energy of the end state, $2|\epsilon_{\text{end}}|$. Its $J_K$ dependence seems to be more complicated and hard to compare, possibly because it involves the removal/addition of two electrons and reflects the total effect of the Kondo interaction on both the end and bulk gaps. A comparison of $\Delta_c$ for $\delta=-0.25\pi$ and $0.75\pi$ suggests that both are reduced by the Kondo interaction, until the local Kondo singlet is formed at the end at $J_K^c$. The subsequent rapid increase of $\Delta_c$ and $\Delta_s$ with increasing $J_K$ indicates that the charge gap is now dominated by breaking the Kondo singlet. $J_K^c$ therefore marks the transition from the topological state to the Kondo state. Combining the above results, we get a phase diagram with the critical Kondo coupling $J_{K}^{c}$ evolving as a function of the phase modulation $\delta$ in $(-0.5\pi,0)$ as plotted in Fig.~\ref{fig5}. We find that the critical Kondo coupling $J_{K}^{c}$ decreases monotonically as $\delta$ approaches zero, possibly due to the reduced energy gap of the end states as shown in Fig.~\ref{fig1}(d). A larger $|\delta|$ tends to enhance the double (empty) occupancy of the conduction electrons at the edges due to the local chemical potential $\pm\mu \sin\delta$, thus disfavoring the Kondo coupling. From the view of the Kondo insulator, the Kondo effect at the edges is weakened by the topological effect, while from the view of the topological insulator, the topological end states can be suppressed by the Kondo effect for sufficiently large $J_K$. The former is consistent with previous understanding of a weakened Kondo effect on the surface of topological Kondo insulators.
  
To summarize, we have studied the interplay of topological and Kondo effects in a one-dimensional Kondo-Heisenberg model with hopping and chemical potential modulation using the density matrix renormalization group method. It is found that the end state has a decouple effect to restrict the formation of Kondo singlet, causing a phase transition from the topological insulating state to the Kondo insulating state at finite critical Kondo coupling $J^c_K$. Our work reveals an interesting interplay of the Kondo physics and the topological physics and may be investigated in real materials, heterostructures, or the optical lattice.

This work was supported by the National Natural Science Foundation of China (Grants No. 12174429 and No. 11974397), and the Strategic Priority Research Program of the Chinese Academy of Sciences (Grant No. XDB33010100).

\end{document}